\documentstyle[epsf]{EuroPhys}
\def\etal{{\hbox{{\tenit\ et al.\/}\tenrm :\ }}}

\def\And{{\rm and\ }}

\def\stars{\bigskip\centerline{***}\medskip}

\newif\ifboo \boofalse

\begin{document}
%
%
%
\newcommand{\taunn}[1]{\begin{picture}(12,#1)
\put(1,6){\circle*{1}} \put(1.5,6){\circle*{1}} \put(2,6){\circle*{1}} 
\put(2.5,6){\circle*{1}} \put(3,6){\circle*{1}} \put(3.5,6){\circle*{1}}
\put(4,6){\circle*{1}} \put(4.5,6){\circle*{1}} \put(5,6){\circle*{1}} 
\put(5.5,6){\circle*{1}} \put(6,6){\circle*{1}} \put(6.5,6){\circle*{1}}
\put(7,6){\circle*{1}} \put(7.5,6){\circle*{1}} \put(8,6){\circle*{1}} 
\put(5,5){\circle*{1}} \put(5,4.5){\circle*{1}} \put(5,4){\circle*{1}}
\put(5,3.5){\circle*{1}} \put(5,3){\circle*{1}} \put(5,2.5){\circle*{1}} 
\put(5,2){\circle*{1}} \put(5.25,1.25){\circle*{1}} \put(5.6,0.6){\circle*{1}}
\put(6.25,0.25){\circle*{1}} \put(7,0){\circle*{1}} \put(7.5,0){\circle*{1}} 
\put(8,0){\circle*{1}} \put(8.5,0){\circle*{1}} \put(9,0){\circle*{1}}
\put(9.5,0){\circle*{1}} \put(10,0){\circle*{1}} \put(10.5,0){\circle*{1}} 
\put(11,0){\circle*{1}}
\end{picture}}
\newcommand{\taurn}[1]{\begin{picture}(12,#1)
\put(11,6){\circle*{1}} \put(10.5,6){\circle*{1}} \put(10,6){\circle*{1}} 
\put(9.5,6){\circle*{1}} \put(9,6){\circle*{1}} \put(8.5,6){\circle*{1}}
\put(8,6){\circle*{1}} \put(7.5,6){\circle*{1}} \put(7,6){\circle*{1}} 
\put(6.5,6){\circle*{1}} \put(6,6){\circle*{1}} \put(5.5,6){\circle*{1}}
\put(5,6){\circle*{1}} \put(4.5,6){\circle*{1}} \put(4,6){\circle*{1}} 
\put(7,5){\circle*{1}} \put(7,4.5){\circle*{1}} \put(7,4){\circle*{1}}
\put(7,3.5){\circle*{1}} \put(7,3){\circle*{1}} \put(7,2.5){\circle*{1}} 
\put(7,2){\circle*{1}} \put(6.75,1.25){\circle*{1}} \put(6.4,0.6){\circle*{1}}
\put(5.75,0.25){\circle*{1}} \put(5,0){\circle*{1}} \put(4.5,0){\circle*{1}} 
\put(4,0){\circle*{1}} \put(3.5,0){\circle*{1}} \put(3,0){\circle*{1}}
\put(2.5,0){\circle*{1}} \put(2,0){\circle*{1}} \put(1.5,0){\circle*{1}} 
\put(1,0){\circle*{1}}
\end{picture}}
\newcommand{\taunr}[1]{\begin{picture}(12,#1)
\put(1,0){\circle*{1}}  \put(1.5,0){\circle*{1}}  \put(2,0){\circle*{1}} 
\put(2.5,0){\circle*{1}}  \put(3,0){\circle*{1}}  \put(3.5,0){\circle*{1}}
\put(4,0){\circle*{1}}  \put(4.5,0){\circle*{1}}  \put(5,0){\circle*{1}} 
\put(5.5,0){\circle*{1}}  \put(6,0){\circle*{1}}  \put(6.5,0){\circle*{1}}
\put(7,0){\circle*{1}}  \put(7.5,0){\circle*{1}}  \put(8,0){\circle*{1}} 
\put(5,1){\circle*{1}}  \put(5,1.5){\circle*{1}}  \put(5,2){\circle*{1}}
\put(5,2.5){\circle*{1}}  \put(5,3){\circle*{1}}  \put(5,3.5){\circle*{1}} 
\put(5,4){\circle*{1}}  \put(5.25,4.75){\circle*{1}}  
\put(5.6,5.4){\circle*{1}}
\put(6.25,5.75){\circle*{1}}  \put(7,6){\circle*{1}}  \put(7.5,6){\circle*{1}} 
\put(8,6){\circle*{1}}  \put(8.5,6){\circle*{1}}  \put(9,6){\circle*{1}}
\put(9.5,6){\circle*{1}}  \put(10,6){\circle*{1}}  \put(10.5,6){\circle*{1}} 
\put(11,6){\circle*{1}}
\end{picture}}
\newcommand{\taurr}[1]{\begin{picture}(12,#1)
\put(11,0){\circle*{1}} \put(10.5,0){\circle*{1}} \put(10,0){\circle*{1}} 
\put(9.5,0){\circle*{1}} \put(9,0){\circle*{1}} \put(8.5,0){\circle*{1}} 
\put(8,0){\circle*{1}} \put(7.5,0){\circle*{1}}
\put(7,0){\circle*{1}} \put(6.5,0){\circle*{1}} \put(6,0){\circle*{1}} 
\put(5.5,0){\circle*{1}}
\put(5,0){\circle*{1}} \put(4.5,0){\circle*{1}} \put(4,0){\circle*{1}}   
\put(7,1){\circle*{1}} \put(7,1.5){\circle*{1}} \put(7,2){\circle*{1}} 
\put(7,2.5){\circle*{1}} \put(7,3){\circle*{1}}
\put(7,3.5){\circle*{1}} \put(7,4){\circle*{1}} \put(6.75,4.75){\circle*{1}} 
\put(6.4,5.4){\circle*{1}}
\put(5.75,5.75){\circle*{1}} \put(5,6){\circle*{1}} \put(4.5,6){\circle*{1}} 
\put(4,6){\circle*{1}} \put(3.5,6){\circle*{1}} \put(3,6){\circle*{1}} 
\put(2.5,6){\circle*{1}} \put(2,6){\circle*{1}}
\put(1.5,6){\circle*{1}} \put(1,6){\circle*{1}}
\end{picture}}
\renewcommand{\thefootnote}{\fnsymbol{footnote}}
%
\euro{43}{6}{664-670}{1998}
\Date{15 September 1998}
\shorttitle{Y. TANAKA \etal  DISCONTINUOUS CRACK FRONTS ETC.}
%
%
%
\title{Discontinuous Crack Fronts of Three-Dimensional Fractures
} 
\author{Yoshimi Tanaka\inst{1}, 
Koji Fukao\inst{2},
Yoshihisa Miyamoto\inst{2}  
\And  Ken Sekimoto\inst{3}
}
\institute{
     \inst{1} Graduate School of Human and Environmental Studies, Kyoto University, \\Kyoto, 606-8501 Japan\\
     \inst{2} Faculty of Integrated Human Studies,
Kyoto University, Kyoto, 606-8501 Japan\\
    \inst{3} Yukawa Institute for Theoretical Physics,
Kyoto University, Kyoto, 606-8502 Japan
}
%
%
\rec{9 September 1997}{in final form 24 July 1998}
%
%
%
\pacs{
\Pacs{46}{30.Nz}{Fracture mechanics, fatigue, and cracks}
\Pacs{61}{41.+e} {Polymers, elastomers, and plastics}
\Pacs{62}{20.Mk}{Fatigue, brittleness, fracture, and crack}
      }
\maketitle
%
%
%

\begin{abstract}
The relation between fracture surface morphology and the three-dimensional
structure of crack fronts is investigated through direct observation 
of brittle cracks in gels. A key notion in this investigation is the 
discontinuity of the 
crack front, whose advancement creates the fracture surface.
We discuss the significance of our findings in the studies of
general three-dimensional brittle fractures.
\end{abstract}
\vspace{-1cm}The crack front of a three-dimensional fracture is a one-dimensional
object whose advancement into a sample creates a crack.
The presence of the third dimension here allows for
phenomena which are not seen in two-dimensional fractures~\cite{recent}.
Actual crack fronts are usually not straight lines;
many step lines are found on fracture surfaces found in rock~\cite{Pollard88}, 
glass~\cite{Sommer69,Abdel77,Doll89,Chevrier95,Beauchamp95}, 
ceramics~\cite{Frechette90}, 
rubber~\cite{Palaniswamy78}, gel~\cite{Tanaka96} (including
edible jelly), etc. Frechette~\cite{Frechette90} has pointed out that a crack 
often becomes an aggregation of crack segments which become interconnected 
during 
the course of fracture propagation and thus that such a crack front does not 
define a  continuous line.
Theoretically, the emergence of non-planar cracks from the front 
of planar cracks has been studied by considering sinusoidally varying 
perturbation  applied to a straight crack front~\cite{Gao92,Xu94} 
and the `planting' of planar crack segments ahead of a parent planar 
crack~\cite{Lawn75,Pollard82,Gell67}.
Little is known, however, about actual fronts of 
well-developed cracks~\cite{Chaudhri86,Marder96}.

Because of the variety of complicated forms exhibited by general
three-dimensional fracture, it would be instructive to begin a study 
of such phenomena by carrying out a geometrical characterization of the most 
elementary structures
of genuine three-dimensional cracks, just as two-dimensional fractures
have been characterized in the terms of elementary mode-$I$, $II$, and $III$ 
cracks.
Mechanical characterization and molecular theory may follow a geometrical
characterization. 
It is the aim of this paper to present the most 
elementary geometrical characterization of three-dimensional cracks,
based on direct observation.
For this purpose, the choices of a suitable material system and investigative 
method are
crucial. Some experiments have been conducted using dense glasses, but in these systems crack growth is very fast, and it is therefore difficult to observe 
crack fronts {\it in situ}.

We have previously reported~\cite{Tanaka96} on pattern formation
involving steps which appear on 
fracture surfaces of gels and on the related dynamical transition with a
critical crack growth velocity. There are several notable advantages of using 
polymer gels for the study of brittle fracture.
First, they exhibit no ductile fracture related to dislocation or similar 
defects around crack fronts. 
In addition, fracture in a gel proceeds very slowly 
compared with that in ordinary solid materials.
Also, from a technical point of view, the optical transparency of gels 
facilitates
{\it in situ} observation of cracks.
In this Letter we report the first direct observations of crack
fronts in well-developed gel fractures. 
(There does exist a related study~\cite{Palaniswamy78},
but this is restricted to the initiation of non-planar cracks from a straight 
crack front.
Discussion related to this point is given below.)
We elucidate the structure of cracks near discontinuities in crack
fronts, and we assert, based on a topological argument, that
this structure is a prototype of general three-dimensional 
brittle fracture.

\begin{figure}
\vspace{-0.2cm}
\epsfxsize=7cm 
\epsfbox{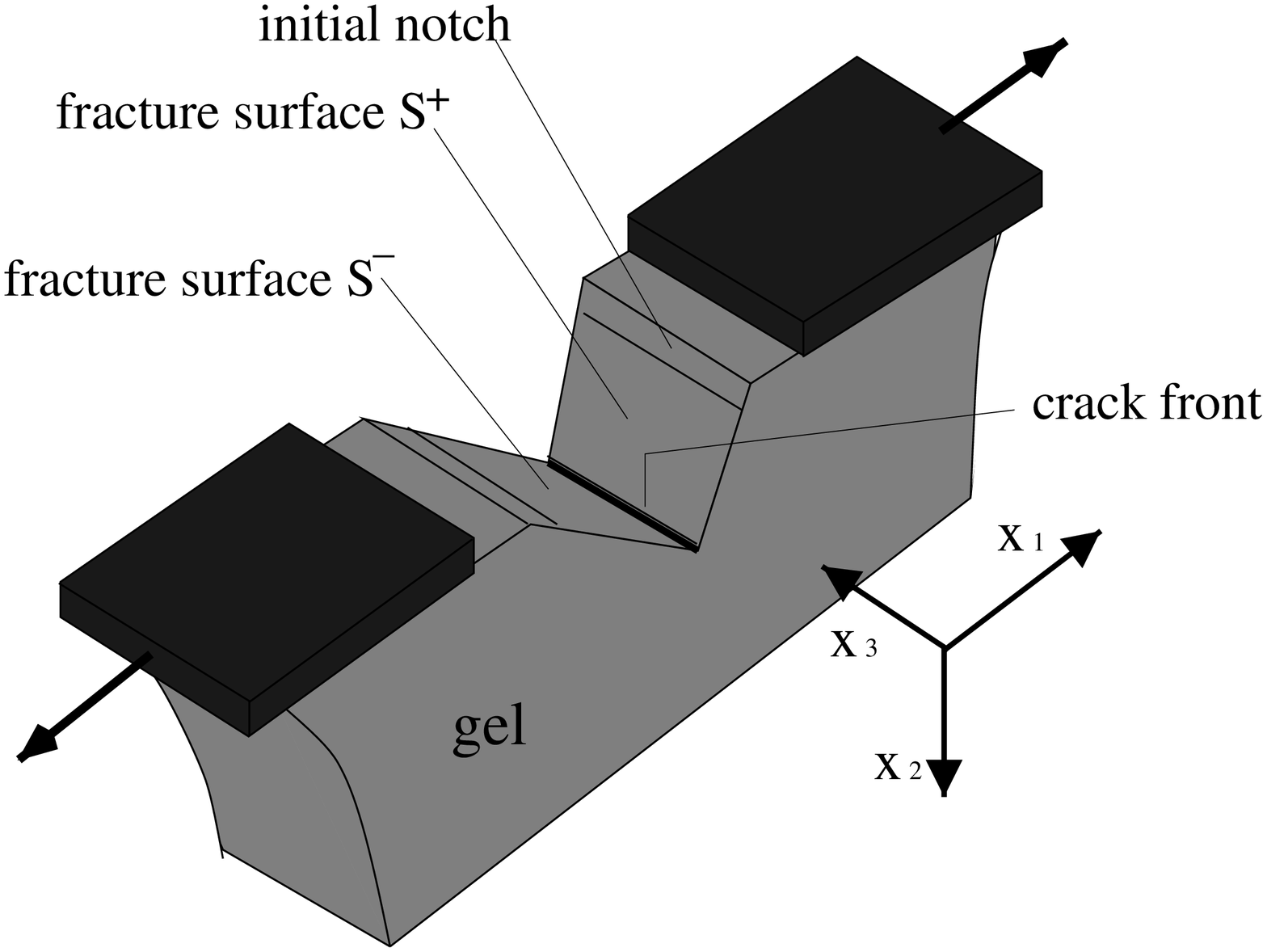}
\begin{picture}(300,10)
 \put(0,5){Fig.1}
 \put(240,5){Fig.2}
\end{picture}
\caption{%
Schematic depiction of a gel sample undergoing fracture.
The Cartesian coordinates are introduced to simplify the discussion 
given in the text.
The sample assumes a pillar-shape of dimension $ 5cm \times 5cm \times 10cm $.
A notch of 3mm depth has been introduced transversally to 
the $ x_1$-direction.  
Fracture is induced by moving apart a pair of the plates which are attached
to opposite sides of the notch.
The crack front propagates in the $ +x_2$-direction.
The crack front is globally parallel to 
the $ x_3$-direction and in the plane of $ x_1=0$.
Behind the crack front a pair of fracture surfaces, $ S^+ $ and $ S^- $, 
are created on the $ x_1>0 $ and $ x_1<0 $ sides, respectively. }
\label{fig1}
\end{figure}

\begin{figure}
\epsfxsize=7cm \epsfysize=4cm
\vspace{-8cm}\hspace*{9.4cm}
\vbox to 4cm{\vfill
{\fbox{See attached Fig.2 (fig\_2.jpg)}}\vfill}
\vspace{3.0cm}
\caption{
Global view of a fracture surface $ S^+ $. 
This photograph was taken after the crack front has passed, from the 
top to the bottom,
through the entire sample. The global speed of the crack front was 0.03 cm/sec.
Step lines are seen as diagonal lines. 
The bar represents a length of 3mm.
}
\label{fig2}
\end{figure}
\begin{figure}

\hspace*{1.2cm}\vspace*{-0.6cm}
\epsfxsize=11cm 
\vbox to 10cm{\vfill\centerline
{\fbox{See attached color Fig.3
(fig\_3.jpg)}}\vfill}
\vspace{-1.0cm}
\caption{
Configuration of a part of a crack near a discontinuity in the
crack front, from which a pair of step lines (one on
$ S^+$ and the other on $ S^-$) extend. 
{\sf a}-{\sf c} show the close-up of the region 
viewed from the $ +x_1<0 $({\sf a}), $ x_1>0 $({\sf b}) and 
$ x_2>0 $  ({\sf c}), sides
{\sf d} is a cross-section of the pair of step lines,
 viewed from the $x_2<0$ side.
The crack was almost closed before taking photographs.
In order to visualize the fracture surfaces, they have been stained by
water-soluble blue paint.
The bars in {\sf    a}-{\sf    c} indicate a length of 0.3mm.
{\sf    e} is the stereo view (parallel-viewing) of the region of a crack where a pair of step 
lines extend.
This structure is determined from {\sf a}-{\sf d}. }
\label{fig3}
\end{figure}

   Our experiment was performed on pillar-shaped gels of dimension 
5cm$\times$5cm$\times$10cm.
The gels were prepared by a standard method using free radical polymerization
of acrylamide (100mM) and bisacrylamide (8.6mM) initiated by
ammonium persulphate and catalyzed by 
tetramethylethylenediamine~\cite{Tanaka78}.
A single notch, 3mm in depth and 0.05mm in thickness, was created along a 
direction
transverse to the longest dimension in each sample by
placing a stainless-steel sheet in the mold used for polymerization.
Samples prepared in this manner were used for the fracture experiments without 
further treatment.
Figure 1 gives a schematic depiction of a gel undergoing fracture, 
where Cartesian coordinates have been
introduced to facilitate the following description.
Two plates were attached to each side of the notch, and 
mechanically driven apart.
As the plates spread, the {\it crack front} emerged from 
 the initial notch and advanced in the direction of increasing $ x_2$.
During the creation of the fracture, the crack front was globally parallel to 
the $ x_3$-direction and in the $ x_1=0$ plane.
Through this process, a pair of fracture surfaces were created behind 
the crack front.
We shall distinguish these fracture surfaces by the symbols $ S^+ $ and $ S^- $,
as shown in Fig.1; that is, $ S^+  $ [ $ S^- $ ]
represents the fracture surface on the $ x_1>0 $ [ $ x_1<0 $] side.
The crack growth rate (i.e., the global speed of the crack front advancing in 
the $ +x_2$-direction in Fig.1) can be controlled through the rate at which 
the two plates are driven apart. In this study we fixed the crack growth rate at 
0.03cm/sec.
(It is known~\cite{Tanaka96} that morphological features of fracture 
surfaces are insensitive to crack growth rates 
below the critical value of dynamical
transition, $v_c \sim 0.1 {\rm cm/sec}$.)

Figure 2 is a photograph of a global view of a fracture surface $ S^+ $,
taken after a crack has expanded across an entire sample.
The many oblique lines on the fracture surface constitute `steps'.
We shall refer to these lines as {\it step lines}.
During the growth of the fracture surfaces $S^+$ and $S^-$ (see Fig.1), 
many step lines extend simultaneously in a pairwise manner so that each 
step line on $S^+$ has a counterpart on $S^-$.
We carefully observed the region near the front of a crack to determine how
a pair of step lines are continuously extended (Figs.3{\sf a}-{\sf d}).
Figures 3{\sf a}-{\sf c} were obtained using the following method.
The growth of the crack was halted at an arbitrary instant, and the crack
surface was then stained with water-soluble paint.
Then, after almost closing the crack, we observed through an optical microscope
the small region where a pair of step lines meet each other.
Figures 3{\sf a}-{\sf c} present, respectively, the views from the $ x_1<0 $ 
(Fig.3{\sf a}), $ x_1>0 $ (Fig.3{\sf b}) and 
$ x_2>0 $ (Fig3.{\sf c}) sides.
Figure 3{\sf d} shows the cross section of such a pair of step lines,
viewed from the $ x_2<0$ side. 
This cross section was obtained by cutting the sample along an $x_2 =$const. 
plane, slightly behind (i.e., corresponding to a smaller $x_2$-coordinate
than) the front of the crack.
In Fig.3{\sf d}, the crack is slightly opened (gap in blue of Fig.3{\sf d}), and 
across it
a pair of the step lines, one on $S^+$ (the upper edges of the gap) and the 
other on $S^-$ (the lower edges), face each other.
This gap has a `tip-to-line' shape, 
which resembles an upside-down the Greek letter ``$\tau$''.
We represent the configuration of crack creating such a gap by the symbol 
$\taunr{10.5}$.

\begin{figure}[b]

\epsfxsize=8cm 

\hspace*{0.4cm}\vspace*{-0.7cm}
\vbox to 8cm{\vfill\centerline
{\fbox{See attached color Fig.4
(fig\_4.jpg)}}\vfill}

\vspace{-0.5cm}
\caption{Classification of the intersections: 
The fracture surface $ S^+ $ around each intersection is schematically
depicted, together with the cross-section of the step lines related to the 
intersection and the microscope image in the case of {\sf a}.
The types of $\tau $-structure entering into or emerging from an 
intersection are identified from these cross-sections. They are
 indicated by the symbols \protect\taurn{8.0} , etc.
These figures show that 
whether the point-like intersections ({\sf    a} and {\sf   b}) or 
the elongated intersections ({\sf    c}-{\sf    e}) are realized 
depends on the types of 
$\tau$-structures entering into the intersections. 
We do not show those classes of intersections which can be obtained 
as mirror images (with respect to the $x_1 x_2$-plane and/or the $x_2 
x_3$-plane)
of the classes shown in {\sf    a}-{\sf    e}.
}
\label{fig4}
\end{figure}

From the above observations, we obtain 
a geometrical characterization of the position at which a pair of step lines
extend (Fig.3{\sf e}).
Here, as in Figs.3{\sf a}-{\sf d}, only a small portion of the sample is shown 
(the frame
in black). The crack, drawn in red and blue, is slightly opened. 
The red curves indicate the {\it crack front}, i.e., the front at which 
the actual irreversible fracture process is taking place.
The blue oblique line which runs to $P_1$ is not the crack front because 
no fracture takes place on it. It is the trace of an end point of the crack 
fronts.
(see below descripion.) 
During the growth of the crack, the crack fronts move obliquely downward, at
some acute angles to both the $x_2$- and $x_3$- axes,
while keeping the shape of the curves unchanged.
As a consequence, the configuration of the fracture surfaces
shown in Fig.3{\sf e} propagates obliquely, from top right to bottom left. 
In Fig.3{\sf e} the crack front consists of
two discontinuous parts; 
the left part extending towards (approximately) the negative $x_3$-direction 
has an endpoint at $P_1$, while the right part extending 
toward the positive $x_3$-direction has an endpoint $P_2$. 
The endpoint $P_2$ is { \it on } the left part of the fracture surface,
which was created by the advancement of the left part of the crack fronts.
We expect this configuration to be quite 
generic, because it is a topological requirement that the crack be 
three-dimensionally connected (see below for further 
discussion).

 By symmetry, there are three other variants of the
configuration shown in Fig.3. Each of these
appears as a mirror image of the form in Fig.3 reflected in the $x_1 x_2$-plane 
and/or the $x_2 x_3$-plane in the coordinate system of that figure.
We will distinguish these four types of configurations by the symbols 
\taunr{10.5}, \taurr{10.5}, \taurn{10.5}, and \taunn{10.5}. 
These symbols are caricatures of the hypothetical cross-sections
of the respective configurations as appearing in planes parallel to the $x_1 
x_3$-plane 
and viewed from the $ x_2<0 $ side, as shown in Fig.3{\sf d}
(see also Figs.4a-e and Fig.5).
We will refer to the portion of a crack 
characterized by these four kinds of configurations as {\it $\tau$-structure} .
When cracks grow under 
global crack-opening forces along the $\pm x_1$-directions, 
all four types of $ \tau$-structure mentioned above 
occur with equal probability.

\begin{figure}
\vspace{-0.2cm}
\epsfxsize=13cm 
\centerline{\epsfbox{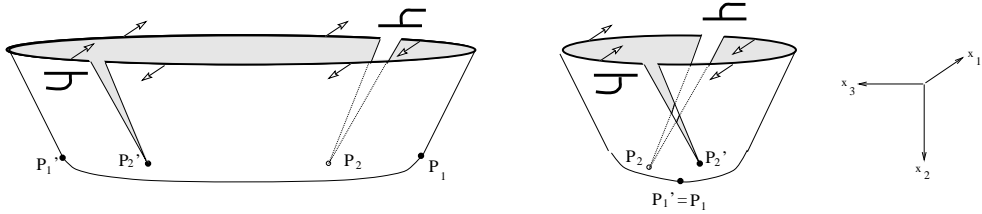}}
\caption{Schematic views of the part of a crack involving two step 
lines of the \protect\taunr{8.0}-type and \protect\taurn{8.0}-type. 
In the top figure, the step lines are separated, where a segment
of the crack front (thick curve) is bounded by the 
points $P_1$ and $P_1^\prime$.
In the right figure, $P_1$ and $P_1^\prime$ merge to form an
elongated intersection (see Fig.4).
}
\label{fig5}
\vspace{-0.4cm}
\end{figure}

An intersection is formed when
a $\tau$-structure of either type \taurr{10.5} or \taurn{10.5} meet with another 
$\tau$-structure of either type \taunn{10.5} or \taunr{10.5}.
Figures 4{\sf a}-{\sf e} represent the intersections of the step lines created 
on a 
crack surface, $ S^+$.
 An intersection which appears as a simple crossing of two straight lines 
(a {\it point-like intersection} for short) occurs if 
the two $ \tau $-structures meeting in this manner 
are combinations of either types  
\taurn{10.5} and \taunn{10.5} (Fig.4{\sf   a} and Fig.4{\sf    b}),
or \taurr{10.5} and \taunr{10.5}.
On the other hand, an intersection that contains a short line element
virtually parallel to the $x_2$-axis (a {\it prolonged intersection} for short)
occurs for a combination of either types \taurr{10.5} and \taunn{10.5} (Fig.4{\sf c}, 
Fig.4{\sf d} and Fig4.{\sf e}), or \taunr{10.5} and \taurn{10.5}. 
In the latter case, real-time observation revealed
the crack growth was temporarily arrested at the point where the
two types of $\tau$-structure met.
Figure 5 schematically displays what should occur for the crack during this 
period;
the point $P_2$ of \taunr{10.5} and its counterpart $P_2^\prime$ 
of \taurn{10.5} are on opposite sides of the opening crack.
As the $x_3$-coordinate of $P_2$ becomes larger than that of $P_2^\prime$  
(the right figure in Fig.5), 
the global mode-{\it I} loading (the arrows in Fig.5) causes 
the `entanglement' of the configurations \taunr{10.5} and \taurn{10.5}.

Figure 4 shows that, given the types of $ \tau $-structure entering an
intersection, the types of $ \tau $-structure 
emerging from the intersection are restricted but not uniquely determined.
In order to solve this selection problem, we would need to know 
the distribution of strain or stress~\cite{Chaudhri86} in the sample. 
The strain or stress near a $\tau$-structure should in turn depend on the 
presence of 
other $ \tau $-structures and intersections, as well as the possible
existence of inhomogeneity in the gel \cite{Bastide}. 
We expect from symmetry considerations that the $x_2 x_3$ component of shear
stress, $T_{23}$, and the $x_1 x_3$ component ($ \it i.e.$ `mode-{\it III}
loading') will affect  the four types of $ \tau 
$-structures differently:
$T_{23}$-stress should differentiate  \taunr{10.5}\, and \,\taunn{10.5}\, from
\,\taurr{10.5}\, and \,\taurn{10.5}, while mode-{\it III} loading should differentiate
\,\taunr{10.5}\, and \,\taurn{10.5} \,from \,\taurr{10.5}\, and \taunn{10.5}.

\begin{figure}
\hspace*{1cm}\vspace{-0.2cm}
\epsfxsize=5cm \epsfysize=2.5cm
\epsfbox{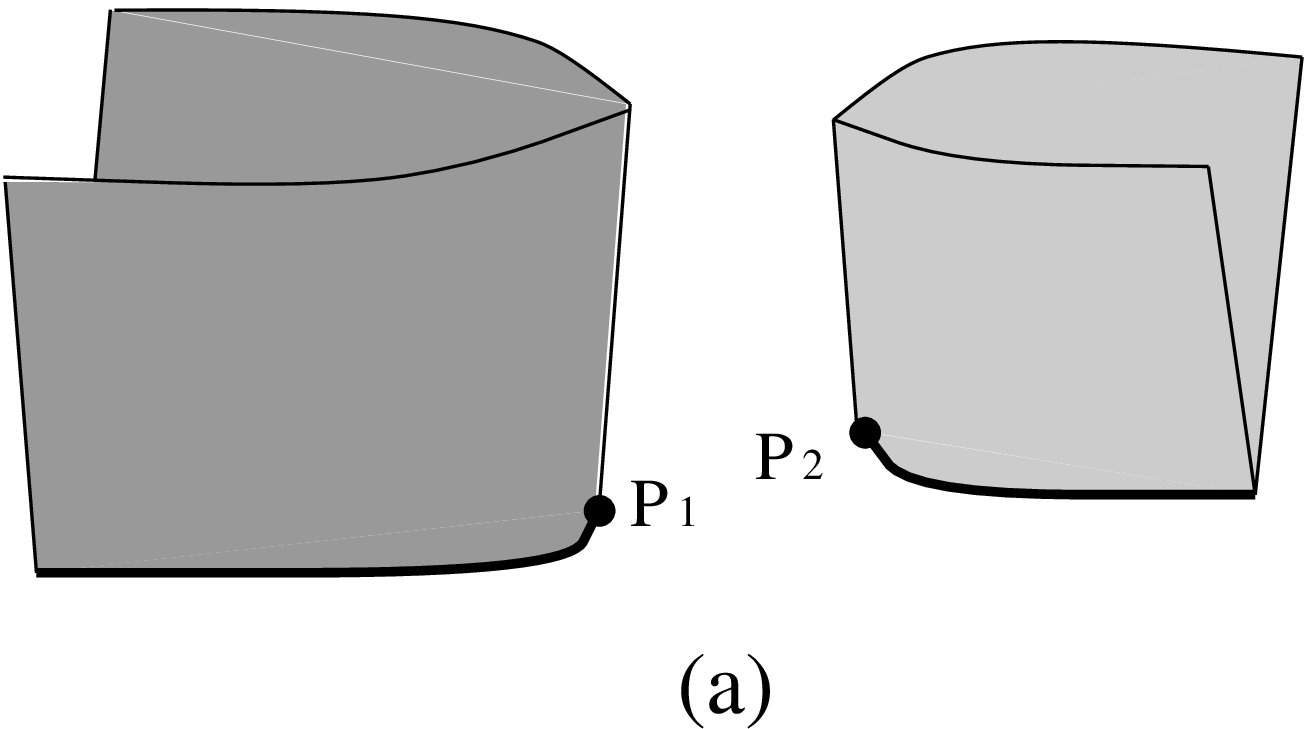}
\hspace*{2.5cm}
\epsfxsize=3.2cm \epsfysize=2.5cm
\epsfbox{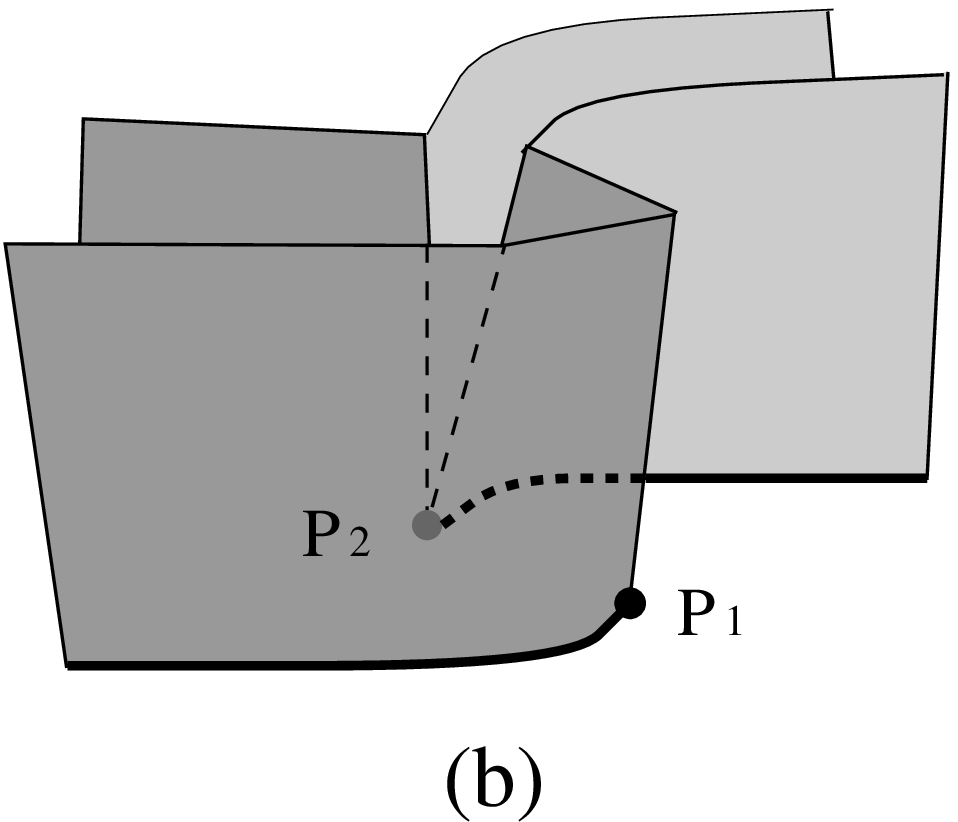}
\caption{
(a) Hypothetical cracks created by a discontinuous crack front. 
These two disconnected cracks would not divide the sample into two pieces. 
(b) The two cracks, being connected, make a continuous crack despite
the presence of a discontinuous crack front. 
The resulting crack structure is equivalent to that 
shown in Fig.3.}
\label{fig6}
\end{figure}

The $\tau$-structures described above should be ubiquitous in the
fracture of three-dimensional materials because the $\tau$-structures embody the 
most generic feature of
discontinuous crack fronts as discussed below.
The most important point in this discussion is that in the process of breaking a 
three-dimensional material into two pieces,
{\it a fracture surface must always be continuous}.
The question we must address is the following: what topologies of cracks
allow for the realization of both the
discontinuity of the crack front and the continuity of the fracture surface?
To answer this in a generic manner, allow us to consider the situation in which, 
starting with two hypothetical partial fracture 
surfaces created by two arbitrary discontinuous crack fronts (Fig.6a),
we connect the partial fracture surfaces in the manner shown in Fig.6b.
The case depicted here corresponds to
the $\tau$-structure \taunr{10.5}, and this structure along with its three variants 
constitute all the generic ways of connecting the two partial fracture 
surfaces shown in Fig.6{\sf a}.

\begin{figure}
\hspace*{1cm}\vspace{-0.2cm}
\epsfxsize=4cm \epsfysize=3cm
\epsfbox{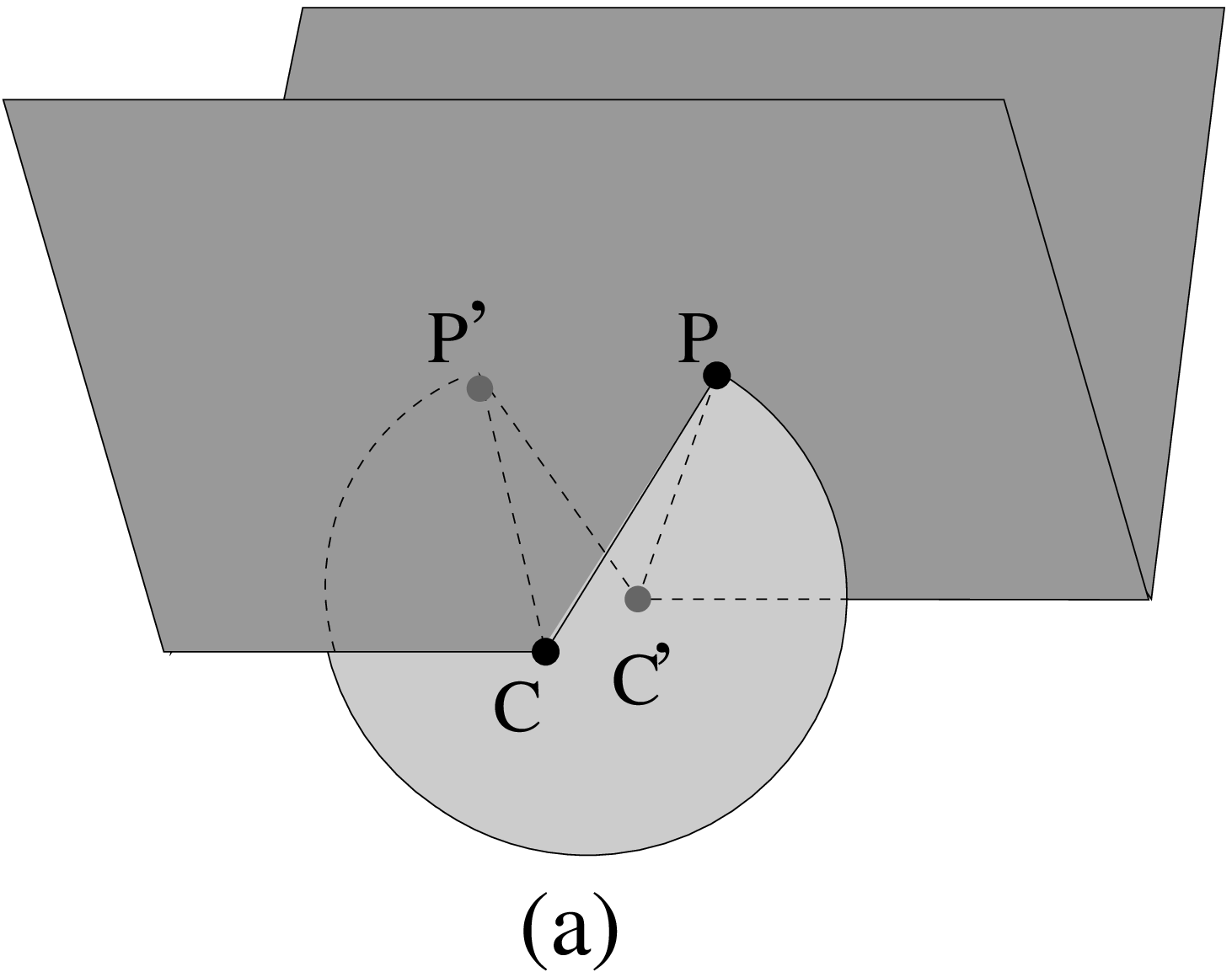}
\hspace*{3cm}
\epsfxsize=4cm \epsfysize=3cm
\epsfbox{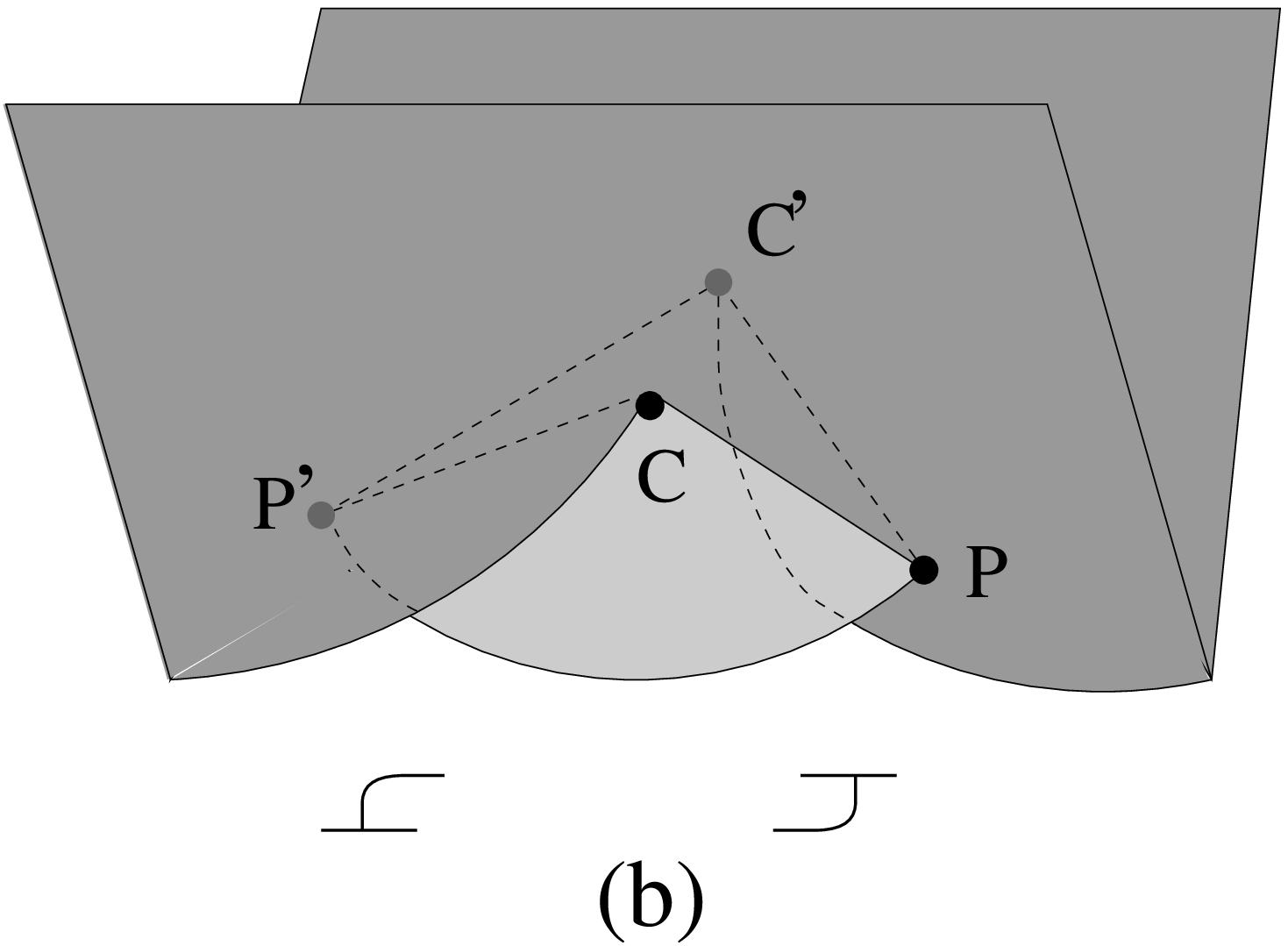}
\caption{
Illustration of the notion of the homology between two types of crack 
structures. 
The structure $(a)$ can be transformed into $(b)$ by a continuous
deformation, pulling down the points
$P$ and $P'$, while raising up the points  $C$ and $C'$: 
$(a)$ is an elementary structure of so-called parahelical cracks
\protect\cite{Palaniswamy78}. 
The dark portion is the parent crack, and the light portion is a 
helical crack, which emerges from the parent crack. 
The points $P$ and $P'$  (the counterpart of  $P_2$ in Figs.1 and 5)
are on opposite sides of the parent fracture surface. 
$(b)$ illustrates the nucleation of a pair of structures, 
\protect\taunr{8.0} and \protect\taurn{8.0}, being juxtaposed in this order.
}
\label{fig7}
\end{figure}
 
We may find as an element the topological object characterized by 
$ \tau $-structure in any fracture surface that consists of more than one part.
For example, in a swollen elastomer~\cite{Palaniswamy78}, an array of
helical crack segments (see Fig.7{\sf a}) appears on the rim
of a parent planar crack (called a
{\it parahelical crack}~\cite{Palaniswamy78}).
Each helical crack can be continuously deformed into the juxtaposition
of \taunr{10.5} \, and \taurn{10.5} (or of \taunn{10.5} and \taurr{10.5}), as shown  
in Fig.7{\sf b}. The points $P$ and $P'$ in Fig.7 correspond to $P_2$ in 
Figs.3 and 5.

Before concluding, we note a polymer physics aspect of the fracture 
of gels. The cooperative diffusion should be induced in the largely deformed
region~\cite{Suzuki97} near a crack front. The length scale 
$ D_{\rm coop }/V_c $,
where $D_{\rm coop}(\sim 10^{-7} {\rm cm}^2/{\rm sec})$
is the cooperative diffusion constant~\cite{DG} and $ V_c(\sim 0.1 {\rm cm})$
is the measured critical crack growth rate of the dynamical 
transition~\cite{Tanaka96}, is just on the order of the blob size $\xi (\sim 100 
\AA$)~\cite{Cohen}.
This suggests macroscopic features of the fracture of gels change depending on
whether or not `diffusion length' $D_{\rm coop}/V $ exceeds the blob size of 
gels.

In this Letter, we have reported the direct observation of crack fronts 
in slow, brittle fracture of gels. 
The structure near the discontinuity of the crack front 
was elucidated, and its implications were discussed from a topological 
viewpoint.
Such a topological viewpoint of three-dimensional fracture should be 
helpful in experimental, numerical
and theoretical studies,
for which a gel can serve as a suitable model system.

\stars
The authors thank Yoshi Oono, Albert Libchaber and an anonymous referee
for critical comments on the earlier version of the manuscript. 
The work was partly supported by a Grant-in-Aid from the Ministry of
Education, Science, Sports and Culture of Japan.



\end{document}